\documentclass[preprint]{aastex63}
\usepackage{multirow}
\usepackage{graphicx}
\usepackage{mwe}
\usepackage[version=4]{mhchem}
\usepackage{xcolor}
\usepackage{xspace}
\usepackage{comment}
\usepackage{placeins}
\usepackage[utf8x]{inputenc}

\received{June 13, 2020}
\revised{August 10, 2020}
\accepted{September 22, 2020}
\submitjournal{ApJ}

\shorttitle{Cosmic ray tracks in astrophysical ices: Modeling with Geant4-DNA}
\shortauthors{Shingledecker et al.}
\graphicspath{{./}{figures/}}

\begin{document}

\title{Cosmic ray tracks in astrophysical ices: Modeling with the Geant4-DNA Monte Carlo Toolkit}

\correspondingauthor{Christopher N. Shingledecker}
\email{cshingledecker@benedictine.edu}

\author[0000-0002-5171-7568]{Christopher N. Shingledecker}
\affil{Department of Physics \& Astronomy, Benedictine College, Atchison, Kansas 66002, USA}
\affil{Center for Astrochemical Studies Max Planck Intitute for Extraterrestrial Physics Garching, Germany}
\affil{Institute for Theoretical Chemistry, University of Stuttgart, Pfaffenwaldring 55, 70569 Stuttgart, Germany}

\author[0000-0002-0619-2053]{Sebastien Incerti}
\affil{Université de Bordeaux, Centre d’Etudes Nucléaires Bordeaux Gradignan (CENBG), Chemin du Solarium, 33175, Gradignan, France}
\affil{CNRS, UMR5797, Centre d’Etudes Nucléaires Bordeaux Gradignan (CENBG), Chemin du Solarium, 33175, Gradignan, France}

\author[0000-0002-1590-1018]{Alexei Ivlev}
\affil{Center for Astrochemical Studies Max Planck Intitute for Extraterrestrial Physics Garching, Germany}

\author[0000-0002-9996-797X]{Dimitris Emfietzoglou}
\affil{Medical Physics Laboratory, University of Ioannina Medical School,  Ioannina GR-45110}

\author[0000-0003-2105-4078]{Ioanna Kyriakou}
\affil{Medical Physics Laboratory, University of Ioannina Medical School,  Ioannina GR-45110}

\author[0000-0003-1684-3355]{Anton Vasyunin}
\affil{Ural Federal University, Ekaterinburg, Russia}
\affil{Visiting Leading Researcher, Engineering Research Institute "Ventspils International Radio Astronomy Centre" of Ventspils University of Applied Sciences, Inženieru 101, Ventspils LV-3601, Latvia}

\author[0000-0003-1481-7911]{Paola Caselli}
\affil{Center for Astrochemical Studies Max Planck Intitute for Extraterrestrial Physics Garching, Germany}



\begin{abstract}
Cosmic rays are ubiquitous in interstellar environments, and their bombardment of dust-grain ice mantles is a possible driver for the formation of complex, even prebiotic molecules. Yet, critical data that are essential for accurate modeling of this phenomenon, such as the average radii of cosmic-ray tracks in amorphous solid water (ASW) remain unconstrained. It is shown that cosmic ray tracks in ASW can be approximated as a cylindrical volume with an average radius that is mostly independent of the initial particle energy. Interactions between energetic ions and both a low-density amorphous (LDA) and high-density amorphous (HDA) ice target are simulated using the Geant4-DNA Monte Carlo toolkit, which allows for tracking secondary electrons down to subexcitation energies in the material. We find the peak track core radii, $r_\mathrm{cyl}$, for LDA and HDA ices to be 9.9 nm and 8.4 nm, respectively - somewhat less than double the value of 5 nm often assumed in astrochemical models.
\end{abstract}



\section{Introduction} \label{sec:introduction}

Within the last decade, a number of observational studies have revealed that cold, prestellar cores are far more chemically complex than has been previously assumed. For example, first \citet{oberg_cold_2010}, and later, e.g., \citet{bacmann_detection_2012}, \citet{cernicharo_discovery_2012}, and \citet{jimenez-serra_spatial_2016} detected a number of species, referred to as Complex Organic Molecules (COMs), in cold cores, including acetaldehyde (\ce{CH3CHO}), dimethyl ether (\ce{CH3OCH3}), methyl formate (\ce{CH3OCHO}), and ketene (\ce{CH2CO}). More recently, \citet{scibelli_prevalence_2020} found, in a survey of 31 starless and prestellar cores in the Taurus Molecular Cloud (TMC), that 70\% contained observable gas-phase abundances of acetaldehyde, and moreover, that methanol (\ce{CH3OH}) was observable towards all sources in their sample. The startling chemical complexity of cold cores was further emphasized by the detection of the aromatic molecule, benzonitrile (\ce{C6H5CN}), in TMC-1 by \citet{mcguire_detection_2018}.

These observational findings are remarkable, in part, because they challenge conventional notions about how such COMs form. It has typically been assumed that COM production occurs mainly within a brief window of time during core collapse in which warming temperatures facilitate the diffusion of radicals on the surfaces of dust-grain ice mantles, as well as the subsequent desorption of COMs, thus produced, into the surrounding gas \citep{garrod_formation_2006,herbst_complex_2009}. However, the observations of COMs at earlier, colder stages of star-formation show that the ability of such species to form at low temperatures has been significantly underestimated, with astrochemical models being only partially successful in shedding light on the underlying formation mechanisms in these regions. For example, the gas-grain code of \citet{vasyunin_reactive_2013} and \citet{vasyunin_formation_2017} was able to qualitatively reproduce the observed abundance of O-bearing COMs by accounting for the increased reactive desorption efficiency on CO-rich ices as well as neutral-neutral reactions efficient at low temperatures. However, this model did not reproduce the observation of N-bearing COMs and overproduced \ce{CH3OH} compared to observations. 

Cosmic rays provide a likely explanation for the aforementioned conundrum. These energetic particles consist mainly of protons with energies of MeV - GeV \citep{indriolo_cosmic-ray_2013}, and are a ubiquitous feature of nearly all astrophysical environments, with the possible exception of protoplanetary-disk midplanes \citep{cleeves_radionuclide_2013}, though recent work by \citet{padovani_cosmic-ray_2018} suggests that cosmic rays might be important there as well. A large body of experimental work has now shown that the interaction between cosmic rays and ices similar to those coating interstellar dust grains can result in both (a) the production and desorption of COMs such as those observed toward cold cores, as well as (b) drive a variety of interface-dynamical mechanisms that can introduce them into the gas (see, e.g,, reviews by \citet{hudson_radiation_2001}, \citet{rothard_modification_2017}, and \citet{arumainayagam_extraterrestrial_2019}). Below, we discuss each of these topics in more detail. 

\subsection{Cosmic-ray driven chemistry}

During the bombardment of some target material (such as a dust-grain ice mantle) by an energetic primary ion (such as a cosmic ray), the primary ion will collide with the atoms and molecules that comprise the material. Following \citet{bohr_theory_1913}, it is customary to divide these collisions into two main categories, namely, inelastic (electronic) and elastic (nuclear) components. For a particle of some energy, $E$, moving through some material of mass density, $\rho$, one can thus describe the energy lost per unit path length ($\mathrm{d}E/\mathrm{d}x$), as

\begin{equation}
    \frac{\mathrm{d}E}{\mathrm{d}x} = \rho(S_\mathrm{e} + S_\mathrm{n}),
    \label{eq:stopping}
\end{equation}

\noindent
where $S_\mathrm{e}$ and $S_\mathrm{n}$ are, respectively, the electronic and nuclear loss functions, with units of cm$^2$ eV/g. The nuclear component $S_\mathrm{n}$ is substantially smaller that $S_\mathrm{e}$ at energies relevant to cosmic rays, and is implicated in changes to the physical structure of the target through, e.g., sputtering and the formation of lattice defect sites, more so than changes to the composition of the target through chemical reactions. Thus, we ignore the elastic component of the primary ion energy loss in this work, and instead focus on the contribution of the inelastic component \citep{sigmund_theory_1969,johnson_irradiation_1991}. 

In general, inelastic collisions between the primary ion and atoms in the material result in the ionization and excitation of target species \citep{spinks_introduction_1990,shingledecker_general_2018}. Ionizing collisions result in the formation of secondary electrons, which have a broad energy spectrum, but the average energies of which do not exceed around 50-70 eV, depending on the target material \citep{spinks_introduction_1990}. Along the trajectories of these secondary electrons, further ionizations and excitations occur. Collectively, the trajectories of the primary ion and all secondary electrons in the target are referred to as the track. Most secondary electrons are stopped near the site of their formation, and thus, the track can approximately be pictured as a cylinder characterized by some radius, $r_\mathrm{cyl}$, which we will refer to as the track ``core.'' 

Within this cylindrical region surrounding the path of the primary ion, the short-lived excited (suprathermal) species drive a rich variety of reactions at even very low temperatures ($<10$ K) that can result in the formation of COMs and even prebiotic molecules \citep{holtom_combined_2005,lafosse_reactivity_2006,hudson_amino_2008}. In \citet{abplanalp_study_2016} it was shown for the first time that reactions involving these suprathermal species are critical for reproducing the chemistry of cold cores, with later investigations showing that their inclusion in astrochemical models results in significant enhancements of the abundance of COMs such as methyl formate under TMC-1 conditions \citep{shingledecker_cosmic-ray-driven_2018}.

The energy deposited in the track core also results in a sudden, sharp increase in the temperature of this region \citep{leger_desorption_1985,bringa_new_2004,ivlev_impulsive_2015}. This rise in temperature further stimulates chemical changes in the target, and in particular, drives reactions with energy barriers that otherwise could not occur at the equilibrium temperature of the ice mantle. Thus, taken together, the combination of suprathermal reactions, and thermal chemistry in the hot track core represent two promising mechanisms that can help explain the observations of COMs in cold prestellar cores. 

\subsection{Cosmic-ray driven desorption mechanisms}

At the interface between the track core and the surrounding vacuum, i.e. the point at which the primary ion enters the target, a hot spot forms in the material with an area of $\sim \pi r_\mathrm{cyl}^2$. In this process, known as impulsive spot heating, the increased temperature of the ice surface within this area significantly increases the rate of thermal desorption. As shown by \citet{leger_desorption_1985}, assuming $r_\mathrm{cyl}<r_\mathrm{grain}$, where $r_\mathrm{grain}$ is the radius of the grain, this process is independent of the actual grain size. 

Conversely, the subsequent process of whole grain heating is not independent of grain size, and occurs as a result of the distribution of the heat deposited in the core throughout the rest of the ice mantle and underlying dust grain. The timescale of this heating has an $a^2$ dependence on the grain radius, $a$, such that for interstellar grains with average radii of $a\approx10^{-5}$ cm, it occurs on the order of nanoseconds \citep{leger_desorption_1985}. Fast, exothermic radical-radical recombinations triggered by such heating could results in the catastrophic loss of the ice mantle through a so-called grain explosion,  first noted by \citet{greenberg_exploding_1973}. 

A separate mechanism that could likewise trigger grain explosions is impulsive spot heating. This process was studied in detail by \citet{ivlev_impulsive_2015} who showed that, depending on (a) the value of $r_\mathrm{cyl}$, and thus, the volume of the core, as well as (b) the amount of energy deposited therein, a dramatic bifurcation in the fate of ice mantles can occur, characterized by the value of a dimensionless parameter referred to as $\lambda$ by Ivlev and coworkers. For $\lambda$ greater than some critical value $\lambda_\mathrm{CR}=9.94$, the bombardment by an energetic ion will similarly result in the sudden loss of the ice mantle via a grain explosion.
 
\subsection{Determining $r_\mathrm{cyl}$}

\begin{figure}[t]
    \centering
    \includegraphics[width=1.0\textwidth]{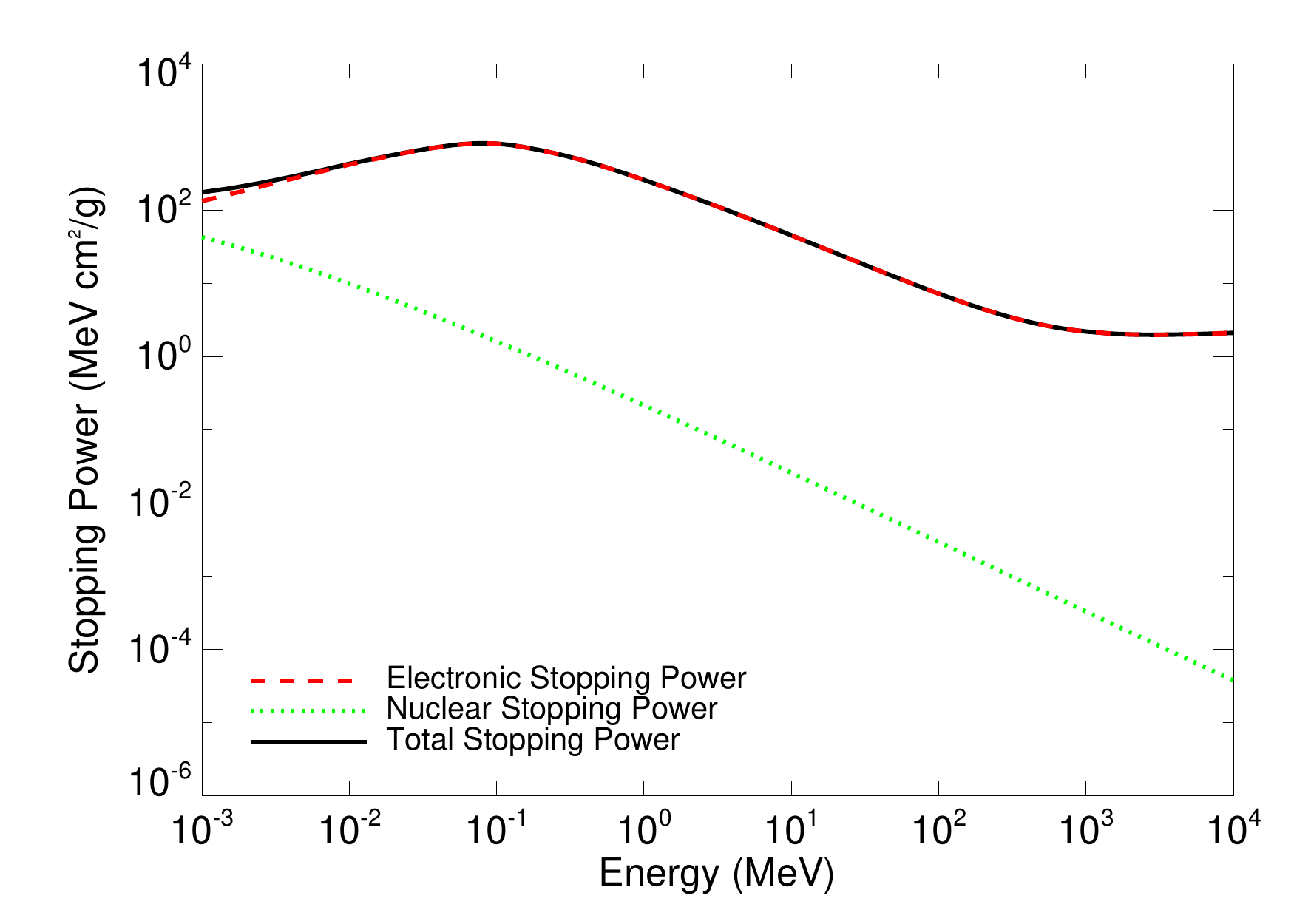}
    \caption{Electronic ($S_\mathrm{e}$), nuclear ($S_\mathrm{n}$), and total ($S_\mathrm{e} + S_\mathrm{n}$) mass stopping power of protons in liquid water, calculated using PSTAR (https://physics.nist.gov/PhysRefData/Star/Text/PSTAR.html)}
    \label{fig:stopping}
\end{figure}

From the preceding discussion, it should hopefully be clear that knowledge of $r_\mathrm{cyl}$ is required for accurate considerations of both the chemistry and interfacial dynamics which occur as a result of the bombardment of a dust-grain ice mantle by a cosmic ray. To date, previous astrochemical works dealing with these topics have typically relied on the estimation from \citet{leger_desorption_1985} of $r_\mathrm{cyl}\approx5$ nm, independent of primary ion velocity \citep{shen_cosmic_2004,ivlev_impulsive_2015,kalvans_temperature_2016}. Conversely, \citet{bringa_electronic_2000} proposed that $r_\mathrm{cyl}$ should \textit{increase} with $\mathrm{d}E/\mathrm{d}x$ due to fast ($<1$ ps) transport of excitation energy over a few lattice spacings. Based on fits to experimental data, they proposed an expression for $r_\mathrm{cyl}$ that has a linear dependence on $\mathrm{d}E/\mathrm{d}x$ and an overall density dependence of $\rho^{1/3}$ \citep{bringa_electronic_2000}.

In principle, however, the radius of the track core qualitatively described by the stopping range of an electron with the average energy of the ejected secondary electrons. Assuming the continuous slowing down approximation, where all particles with the same energy are assumed to travel the same average distance \citep{johnson_energetic_1990} this range for secondary electrons with an average energy $W_\mathrm{av}$  is given by

\begin{equation}
    r_\mathrm{cyl} \approx \int_0^{W_\mathrm{av}} \left(\frac{\mathrm{d}E}{\mathrm{d}x}\right)_\mathrm{electron}^{-1}\mathrm{d}E = \frac{1}{\rho}\int_0^{W_\mathrm{av}} \left(\frac{1}{S_\mathrm{e, electron}}\right)\mathrm{d}E.
    \label{eq:csda_se}
\end{equation}

\noindent
Here, $\left(\mathrm{d}E/\mathrm{d}x\right)_\mathrm{electron}$ is the energy deposited per unit path length for electrons, and similarly, $S_\mathrm{e, electron}$ is the electronic loss function, also for electrons. In principal, one could use Eq. \eqref{eq:csda_se} to estimate the track core radius, however such an approach would require accurate analytical expressions for the electronic stopping losses, the derivation of which are beyond the scope of the current work. Nevertheless, Eq. \eqref{eq:csda_se} is still useful, since it allows us to qualitatively understand the results obtained using the Monte Carlo methods utilized here, for example, by comparing the dependence of our results on material density with the $1/\rho$ dependence one would expect from Eq. \eqref{eq:csda_se}.

Thus, we are presented with three conflicting predictions as to the value of $r_\mathrm{cyl}$. In this work, we seek to resolve this confusion and establish more explicitly the value of this critical parameter based on a leading-edge Monte Carlo code, designed to yield astrochemically relevant values for amorphous solid water over a range of cosmic ray proton energies. The rest of this work is arranged as follows: in \S\ref{sec:methods} we provide details of our model and computational approach, in \S\ref{sec:results} we present the results of calculations and discuss their astrophysical significance, and finally, in \S\ref{sec:conclusions} we summarize our conclusions. 

\FloatBarrier
\section{Methods} \label{sec:methods}

\subsection{Geant4-DNA} \label{sec:g4dna}

For this work, we have employed the GEANT4 v10.6 Monte Carlo simulation toolkit \citep{agostinelli_geant4simulation_2003,apostolakis_geometry_2009,allison_recent_2016}, which was initially designed to simulate systems relevant for high-energy physics. However, the flexibility of the code allows for its application to problems in a wide variety of fields, including medical physics and astrophysics. The toolkit was later extended by the Geant4-DNA project to simulate microdosimetry through the addition of additional physical processes, such as excitation and charge exchange, that were not included in the original Geant4 code, and allow for the accurate modeling of collisional events down to energies of a few eV \citep{incerti_geant4-dna_2018,bernal_track_2015,incerti_comparison_2010,incerti_geant4-dna_2010}. The original motivation for this extension, as indicated by the addition of ``DNA'' to the name of the toolkit, was to investigate the effects of ionizing radiation in biological systems, including especially DNA and RNA damage.  

For this study, we utilized the \texttt{G4EmDNAPhysics\_option2} physics list. A full description of the processes and valid energy ranges for particles considered in Geant4-DNA can be found in \citet{incerti_geant4-dna_2018}, and includes elastic electron scattering, shell ionization cross sections (5 shells), excitation cross sections (5 levels), full secondary electron cascade generation from individual shells (using shell-specific differential ionization cross sections), vibrational excitation, and molecular attachment. In the context of Geant4-DNA, and indeed, of all similar MC codes the track is defined as the collection of the above-mentioned interaction ``points,'' which occur at a given set of x,y, and z coordinates in our simulated volume. Among other restrictions of quantum origin, the spatial extent of the ``point'' cannot be smaller than the dimensions of the target molecule which, in the case of water, is governed by the $\sim0.3$ nm diameter of the molecule.

\subsection{Simulation of Ice Bombardment}

In the interstellar medium, water ice forms on dust grains through the adsorption and subsequent reaction of, e.g., O, H, OH, and \ce{O2} \citep{ioppolo_water_2010,cuppen_water_2010}. This ice exists mostly as amorphous solid water (ASW). The properties of this glassy metastable material depend on the physical conditions under which it formed, combined with the effects resulting from any changes of these conditions and of any subsequent processing. The two main types of ASW of astrophysical relevance are low-density amorphous (LDA) and high-density amorphous (HDA) ice, which have densities of 0.94 and 1.1 g cm$^{-3}$, respectively \citep{narten_diffraction_1976,jenniskens_structural_1994}. In the ISM, the bombardment of LDA by energetic particles similar to cosmic rays has been found to result in its compactification, leading possible to HDA \citep{palumbo_morphology_2005,mitterdorfer_small-angle_2014}. 

By default, the Geant4-DNA simulations include data only for liquid water, however, given the structural similarity of it with the glassy ASW, we here approximate ASW by scaling the density of the material in our model. Since the mean-free-path of electrons is dependent on the density of the material, we here perform calculations at densities relevant to both LDA and HDA. In the context of MC transport of energetic charged particles, several studies have examined the energy-loss function properties of solid water (amorphous and hexagonal ice) and found to be very similar to that of liquid water \citep{emfietzoglou_comparison_2007,emfietzoglou_consistent_2007,emfietzoglou_semi-empirical_2008}. Therefore, we expect the energy loss of charged particles in ASW and liquid water to be quite similar and as such, the inclusion of ASW-specific interaction cross sections will not appreciably change the results or conclusions described in the following sections. Perhaps the main uncertainty in the present work comes from the lack of rigorous corrections to the first Born approximation for inelastic electron scattering below about 100 eV. Geant4-DNA has already implemented such corrections in some of its physics models, including the one used in this work, but they are mostly phenomenological. It is possible that these uncertainties may influence very low energy electron transport at the few nm scale, however, an investigation in this matter is beyond the scope of our study.

In our simulations, we represent the ice as a cube with edges 1 $\mu$m in length. This ice is then bombarded with protons - the major constituent of cosmic rays - with energies between 100 keV and 100 MeV, which covers both the peak and subsequent fall-off of the electronic stopping power, as depicted in Fig. \ref{fig:stopping}. Incident primary ions are assumed to collide with the ice normal to the surface in the center of the topmost side. Since the stopping length of protons in the energy range considered here is larger than the 1 $\mu$m thickness of our ice, they are able to pass completely through, at which point they are considered to have left the system and are not followed any further. For each model run, the simulation begins with the first collision of the primary ion, and ends when all secondary electrons reach energies of around 7.4 eV, below which Geant4-DNA does not currently simulate them. 

Finally, in order to aid in the calculation of track core widths, we have disabled elastic scattering processes for the incident protons\footnote{proton\_G4DNAElastic} as well as neutral hydrogen\footnote{hydrogen\_G4DNAElastic}, since the incident \ce{H+} can capture an electron from the target material. Since, as one can see from Fig. \ref{fig:stopping}, the nuclear component of the stopping power is more than $\sim2$ orders of magnitude less than the electronic component, and moreover, since it does not effect the width of the track core, this assumption should not hinder the accuracy of the resulting calculated values of $r_\mathrm{cyl}$. 

\FloatBarrier
\section{Results \& Discussion} \label{sec:results}

For each initial proton energy in the considered energy range of 0.1 - 100 MeV, 1000 simulations were performed, with different randomly-chosen seeds used for each model run. In order to aid in the determination of track core radii, each three-dimensional track was projected onto a (61 nm)$^2$ 2-D surface in the $y$-$z$ plane of the simulated volume (with the $x$ coordinate here giving the vertical component), represented as a grid of $601\times601$ bins, chosen to encompass the entirety of the track core. For the total track, each bin in our grid was assigned a value equal to the total number of collisions with $y$-$z$ coordinates within the area covered by the bin, averaged over the 1000 model runs. 

One advantage of Geant4-DNA is that the type of each collision simulated in our model is recorded, as well as the resulting energy loss from the primary ion or any other particle generated (e.g. secondary electrons). These data were used to estimate the following four radii:

\begin{itemize}
    \item Total Ionizations ($r_\mathrm{ion}$): Here, we mean the subset of inelastic collisions which result in the ionization of a bulk species and the concomitant formation of a secondary electron. Electrons formed via collisions of bulk species with the incident primary ion are referred to as first-generation secondary electrons, and these, in turn, can be formed with sufficient energy to ionize yet more bulk species, thereby forming second-generation secondary electrons, which, depending on their energies, can generate still higher-generation secondaries. As noted in \S\ref{sec:g4dna}, all such electrons are followed in the code until they reach energies of $\sim7.4$ eV.
    \item Total Excitations ($r_\mathrm{exc}$): Here, we mean the subset of inelastic collisional processes which result in the excitation of a bulk species to some higher bound state. In the code, these collisions can involve either first-generation secondary electrons, or any higher generation of secondary electron. A brief description of the inelastic collisional processes considered in our code is given in \S\ref{sec:g4dna} and the references mentioned there.
    \item Total Collisions ($r_\mathrm{tot}$): Here, we mean all collisions, both elastic as well as inelastic, e.g. ionization and excitation, by all generations of electrons. This representation of the track is interesting, since it most closely corresponds to the ``real'' track, as described in \S\ref{sec:g4dna}.
    \item Average Energy Deposition ($r_\mathrm{energy}$): In each inelastic collision, some amount of energy is lost by either the primary ion or secondary electron, and these energy losses are explicitly recorded as outputs of a Geant4-DNA simulation. By thus calculating the average total energy deposited in collisions occurring with $y$-$z$ coordinates covered by each 1 \AA$^2$ bin in our grid, we obtain energy deposition maps from which we can calculate track core radii as described below.
\end{itemize}

\begin{figure}[htb]
\centering
\includegraphics[width=0.5\textwidth]{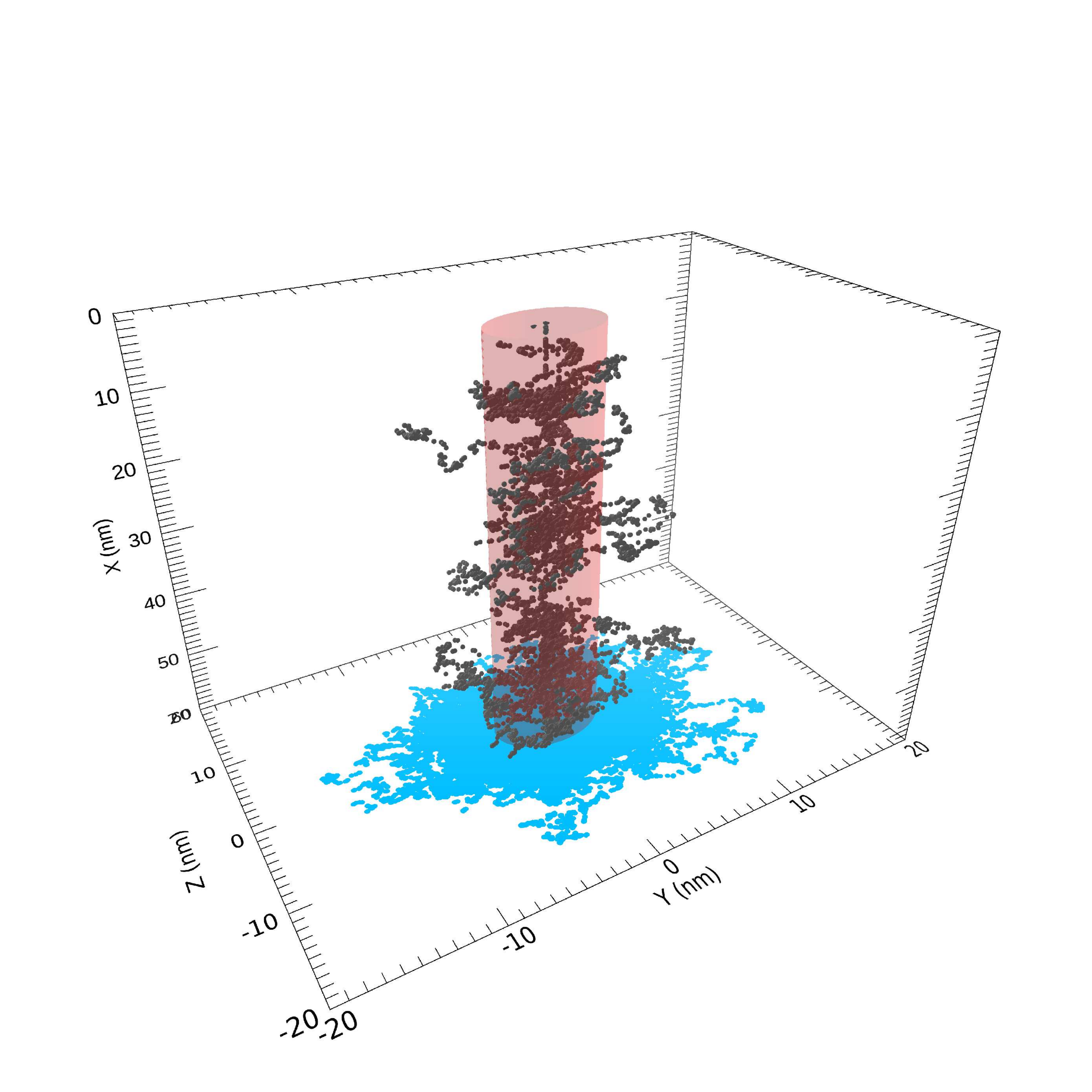}
\caption{A portion of the track for a 0.1 MeV proton, shown in grey, with the cylindrical track-core region depicted in red and the projection onto the $y$-$z$ plane shown in light blue. Note: here, the incident ion enters the ice through the center of the top surface.
\label{fig:3d}}
\end{figure}

These radii were determined by calculating the cumulative distribution of the averaged values for circles of radii from 0.1 - 30.0 nm from the origin, i.e. where the primary ion enters the ice, in steps of 1 \AA. Based on these cumulative distribution maps the track core was defined as the circular region inside of which $1\sigma$ (68.27\%) of the events occurred. This value was chosen because it was found to best approximate the fairly stable track core region across the range of energies considered, unlike the much larger $2\sigma$ (95.45\%) or $3\sigma$ (99.73\%) radii, which better traced the less numerous secondary electrons of the track ``penumbra.'' Note that in the following discussion we assume that $r_\mathrm{cyl} = r_\mathrm{tot}$. To show the three-dimensional structure of a track, as well as to illustrate a magnified view of our total 1 $\mu$m$^3$ volume, we show in Fig. \ref{fig:3d} a portion of the track of a 0.1 MeV proton in LDA ice, along with both the projection onto the $y$-$z$ plane and the cylindrical, 1$\sigma$ track-core region. We note that, when averaged over 1000 simulation runs, the irregularities formed by the individual secondary electron paths visible in the projection onto the $y$-$z$ plane in Fig. \ref{fig:3d} are averaged out, and an approximate radial symmetry emerges.

\subsection{LDA Results} \label{sec:lda_results}

The results of our simulations for LDA ice are listed in Table \ref{tab:rlda}. There, in addition to listing the total radius counting all collisions along the track ($r_\mathrm{tot}^{LDA}$), we further list radii in which only ionization or excitation collisions of the total track were counted. These radii  are plotted as a function of initial particle energy in the left panel of Fig. \ref{fig:radii}. In addition, In Fig. \ref{fig:ldaset} of Appendix A, we show the 1- and 2D representations of the total track, as well as the several sub-tracks considered here. To better illustrate our method of obtaining the values given in Table\ref{tab:rlda}, we show in Fig. \ref{fig:cdf_lda} a representative example of the cumulative distribution plots for a 0.1 MeV \ce{H+}, as well as contours indicating the track core radii.

From Fig. \ref{fig:radii}, one can see that the calculated radii for ionization, $r_\mathrm{ion}$, are the smallest among those shown. This result is expected since secondary electrons only have the energy to undergo $\sim$ a few ionization collisions before falling below the ionization threshold of the material. Moreover, ionization cross-sections are generally on the order of $\sim$1 order of magnitude larger than excitation cross-sections \citep{johnson_irradiation_1991}, and therefore, the former are expected to occur with a similarly higher probability than the latter \citep{shingledecker_new_2017}. Thus, for secondary energies above the ionization threshold of the material, ionizing collisions will dominate, and thus, will occur within a shorter distance to the point of formation than for excitation collisions.

Since we track the average energy deposited per primary ion, rather than the total energy deposition, the fact that more energy is lost in ionizing collisions (being equal to the ionization energy of the material) than excitation collisions ($<10$ eV) means that $r_\mathrm{energy}$, closely follows $r_\mathrm{ion}$, though at a slightly larger value due to the contribution of energy deposited during excitation collisions. 

One can also see that the total radius of the track core, $r_\mathrm{tot}$, is nearly identical to the radius of excitation collisions, $r_\mathrm{exc}$, from 0.1 - 100 MeV. Over this energy range, these radii display values of $\sim9.0\pm1.0$ nm. The close association between the total and excitation radii is again not surprising, since, once falling below the ionization threshold, excitation collisions of various kinds, e.g. electronic and vibrational, are the dominant inelastic processes until the electron falls below the excitation energy threshold of the material.

These radii characterize the behaviour of secondary electrons in the material and are thus sensitive the their energies upon formation. As noted by \citet{rudd_electron_1992}, for an incident proton with energy $E_\mathrm{ion}$, momentum, $p$, and mass $m_\mathrm{p}$, the maximum secondary electron energy, $W_\mathrm{max}$ which can be produced by that ion is given by

\begin{equation}
    W_\mathrm{max}(p) = 2m_e v_0^2 = 4\left(\frac{m_\mathrm{e}}{m_\mathrm{p}}\right)E_\mathrm{ion},
    \label{wmax}
\end{equation}

\noindent
where $m_e$ and $v_0$ are, respectively, the electron mass and incident ion velocity. Thus, the maximum secondary electron energy capable of being produced by a 0.1 MeV proton is 218 eV, while for a 100 MeV proton the value is 218 keV. However, the probability of producing such a secondary electron with energy $W_\mathrm{max}$ is small, with the majority of secondary electrons having a broad energy spectrum but with average energies of not more than 50-70 eV, depending on the target material. The secondary electron energies indicated by our radii do not show the linear dependence on $E_\mathrm{ion}$ implied by Eq. \eqref{wmax}. In fact, our results are in agreement with semi-empirical estimations of secondary electron energies by \citet{rudd_differential_1988}, who found that the average secondary electron energy, $W_\mathrm{av}$, increased until it saturates at around incident proton energies of a few tenths to a few MeV, again depending on the target (cf. Fig. 10 of \citet{rudd_differential_1988}). The results presented in Fig. 10 of \citet{rudd_differential_1988}, along with the expression for $r_\mathrm{cyl}$ given in Eq. \eqref{eq:csda_se}, together provide a very reasonable description of the energy dependence of the radii shown in Fig. \ref{fig:radii}, where the track core radii will likewise follow changes in $W_\mathrm{av}$. This ability to qualitatively describe the findings presented in Fig. \ref{fig:radii} in light of Eq. \eqref{eq:csda_se} and the work of \citet{rudd_differential_1988} is thus a reassuring indicator of the reliability of our approach.

We note that the relationship between our derived $r_\mathrm{cyl}$ values and incident proton energies, though in agreement with the expected average secondary electron energies calculated by \citet{rudd_differential_1988}, is not what one would expect based on the formula derived by \citet{bringa_electronic_2000}, which is linearly dependent of the stopping power $(\mathrm{d}E/\mathrm{d}x)$. As shown in Fig. \ref{fig:stopping}, the electronic stopping power varies by around 3 orders of magnitude over the 0.1 - 100 MeV range considered here; however, as shown in Fig. \ref{fig:radii}, our track core radii vary by only a factor of a few over that range.

\subsection{HDA Results} \label{sec:hda_results}

\begin{figure}[htb]
\plottwo{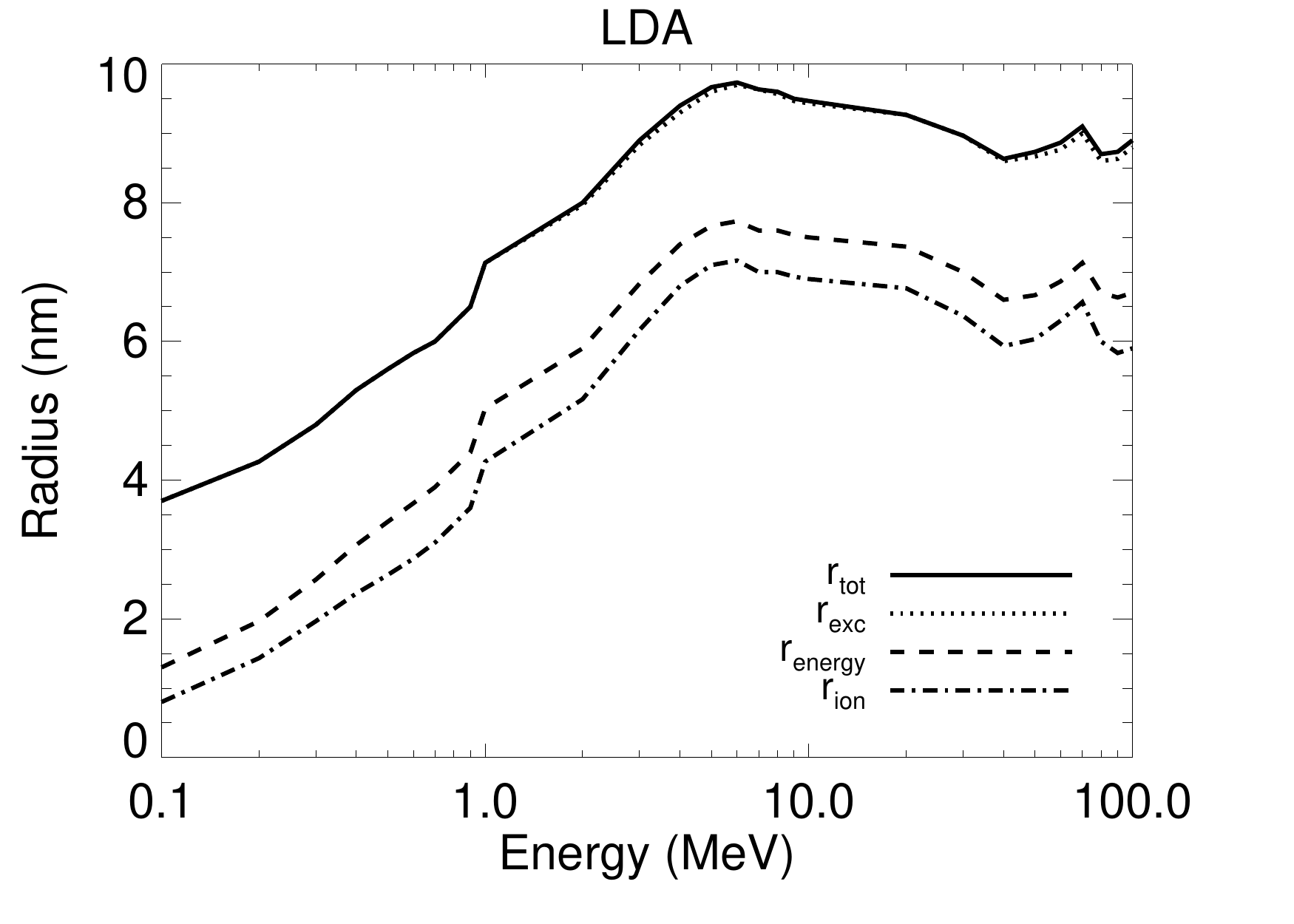}{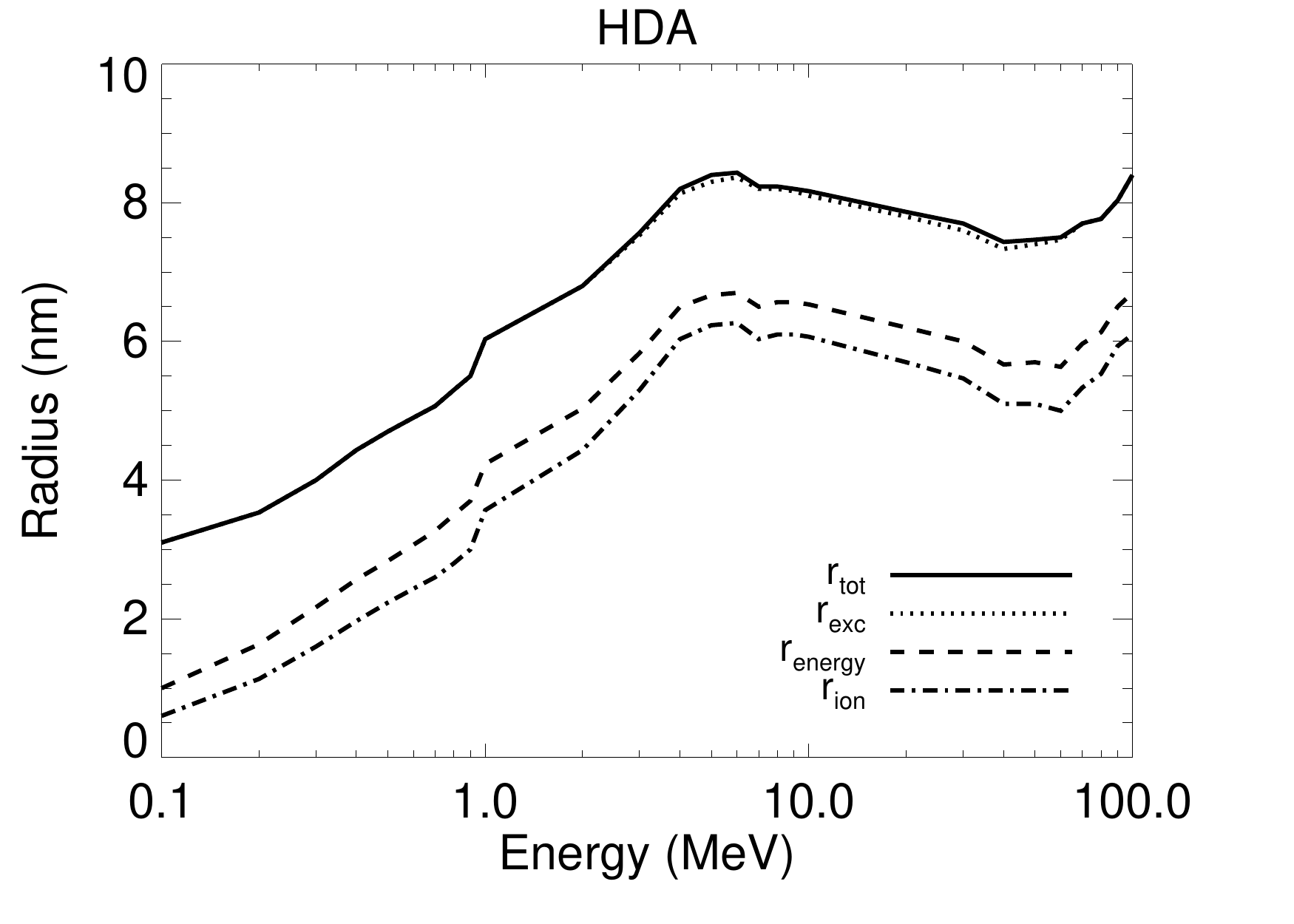}
\caption{Radii calculated with Geant4-DNA for both LDA (left) and HDA (right) ices.
\label{fig:radii}}
\end{figure}

\begin{figure}
\gridline{
          \fig{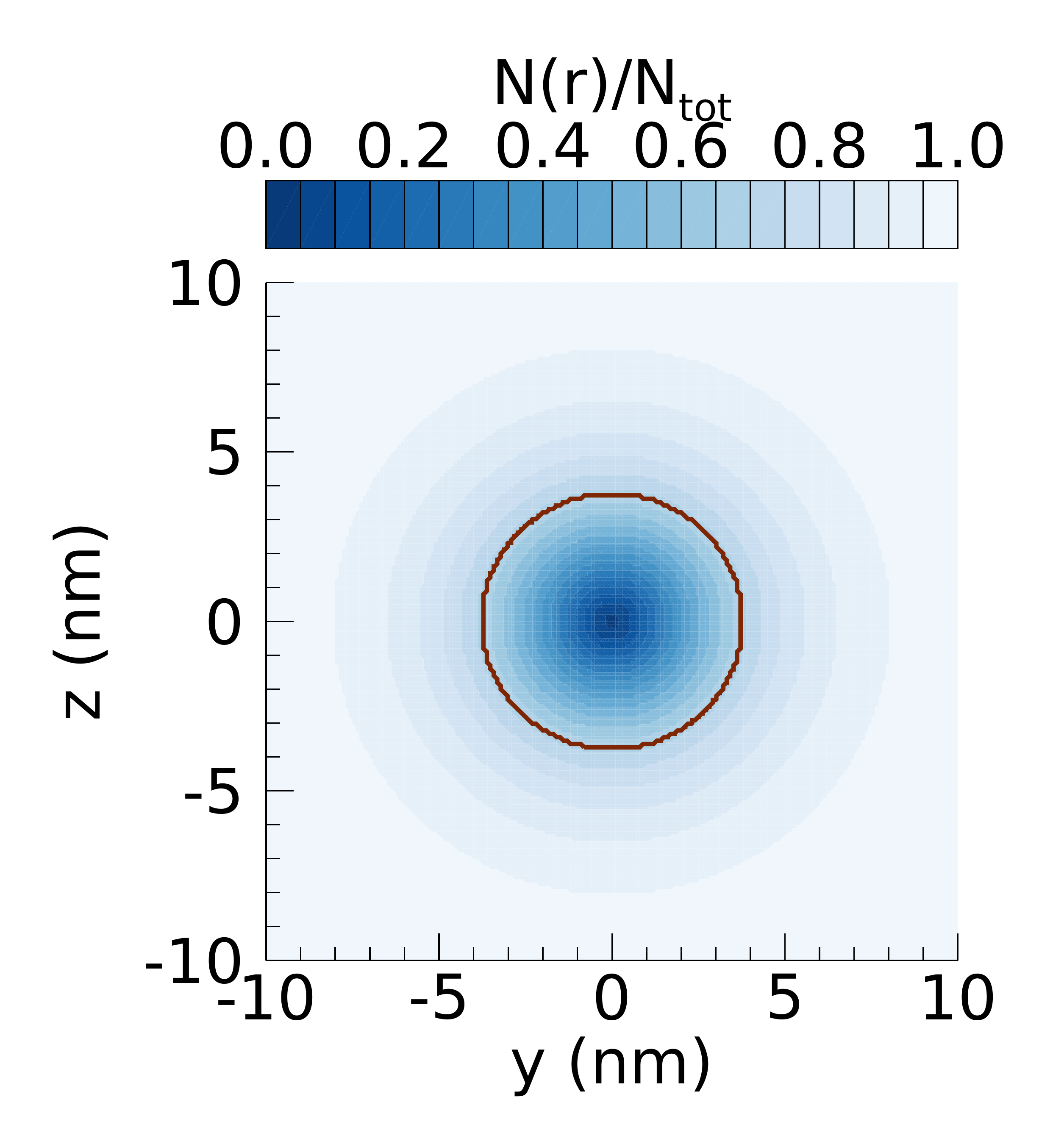}{0.48\textwidth}{(a)}
          \fig{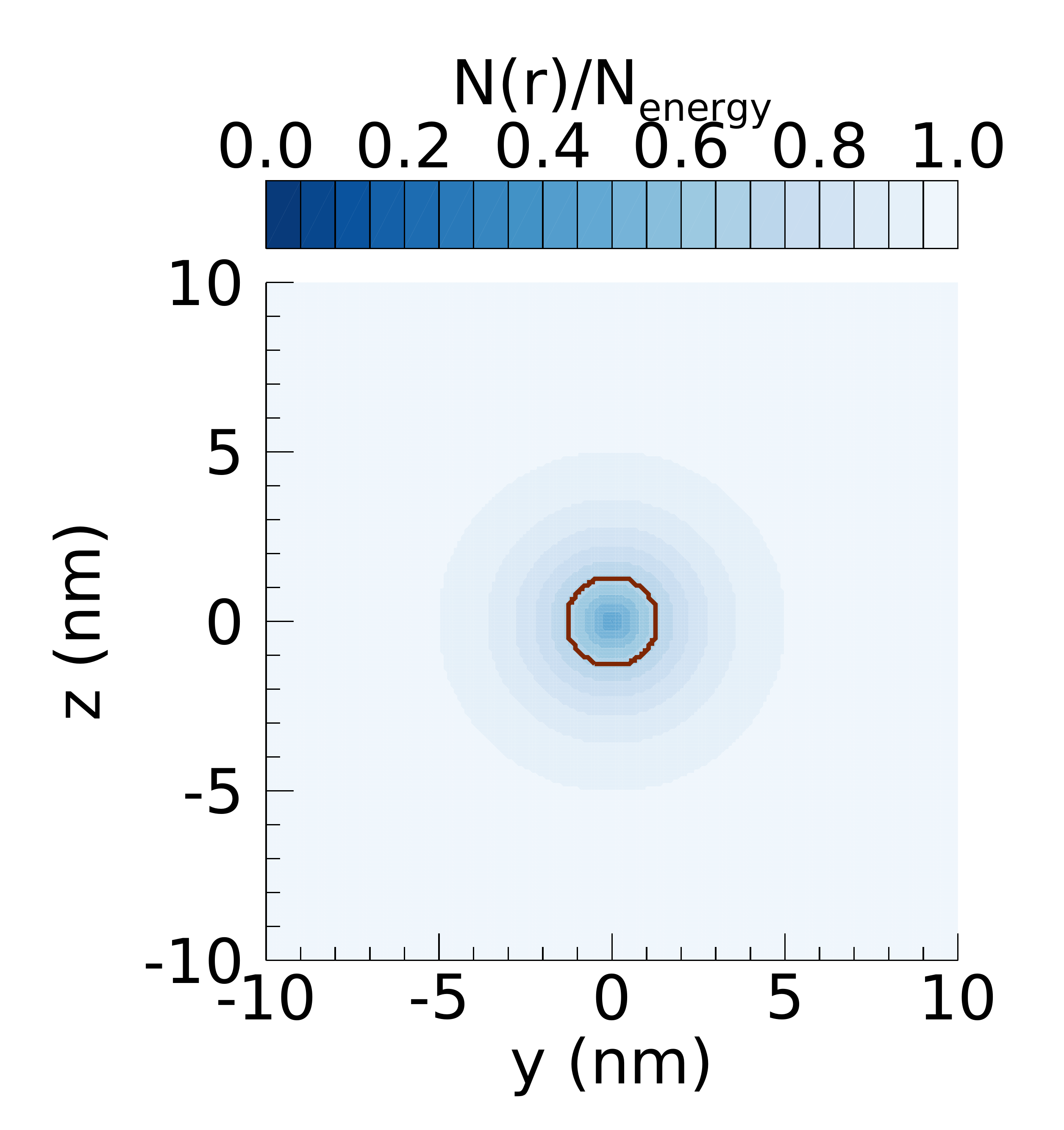}{0.48\textwidth}{(b)}
          }
\gridline{
          \fig{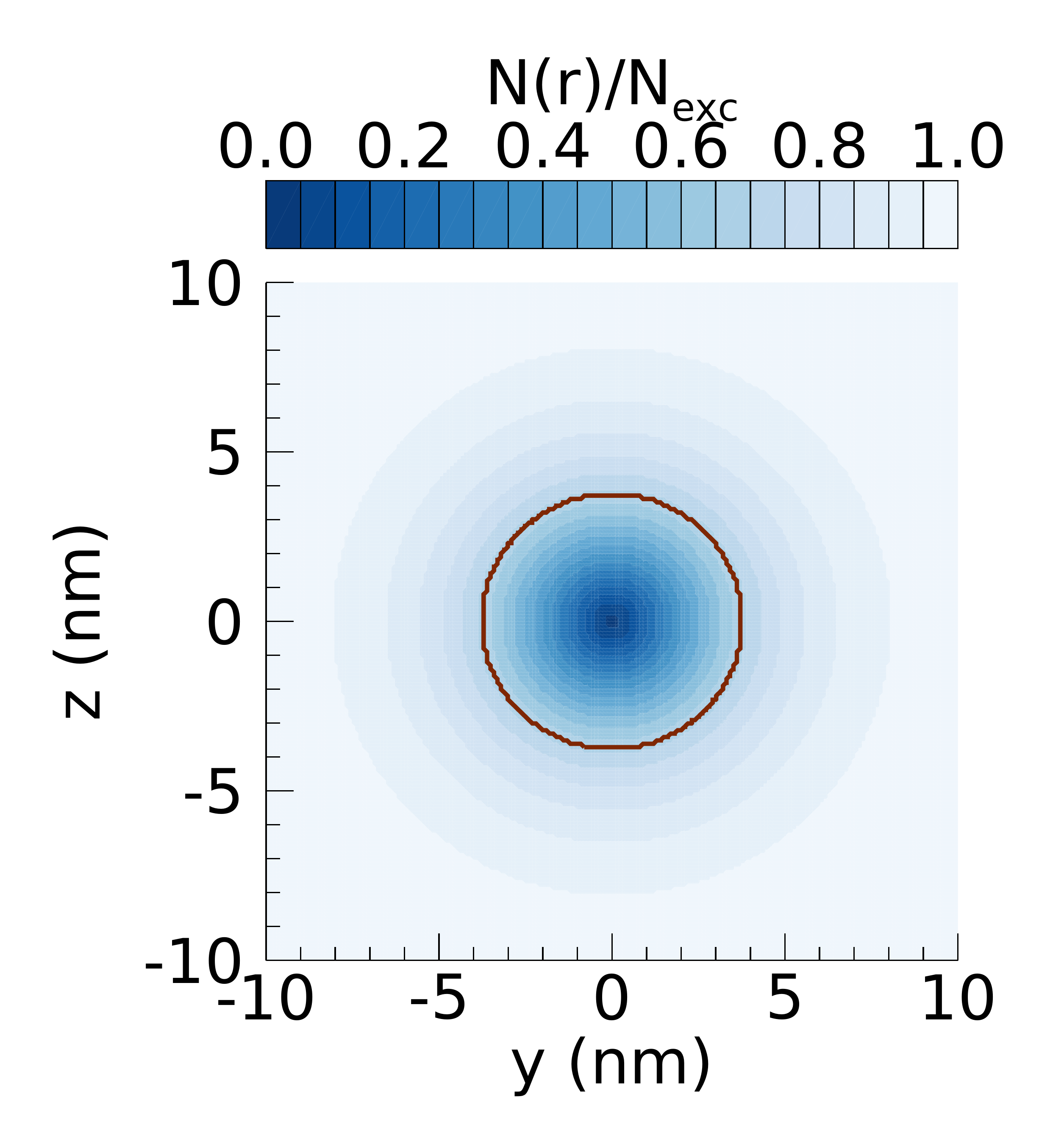}{0.48\textwidth}{(c)}
          \fig{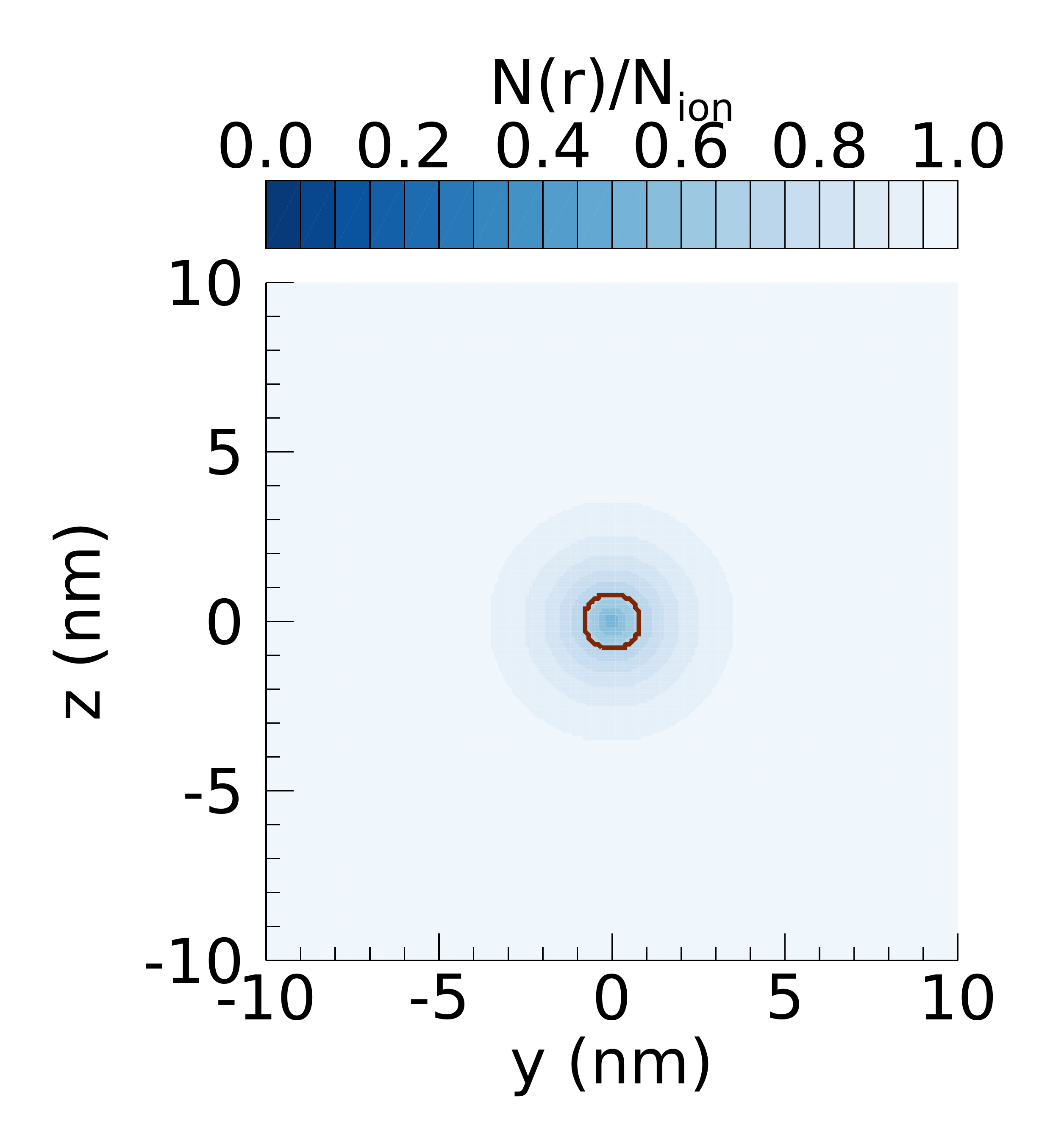}{0.48\textwidth}{(d)}
          }
\caption{The average number of total collisions (a), energy deposition (b), excitation collisions (c), and ionization collisions (d) for a 0.1 MeV \ce{H+} in LDA ice at radius $r$, $N(r)$, over the total value, $N(tot)$. The radius at which $1\sigma$ ($\sim68$\%) of the collisions occur is represented by a red circle. \label{fig:cdf_lda}}
\end{figure}

Calculated radii for high-density amorphous (HDA) ice are also listed in Table \ref{tab:rlda}, and plotted as a function of energy in the right panel of Fig. \ref{fig:radii}. The 1- and 2D representations of the total track, as well as the several sub-tracks considered here, are given in Fig. \ref{fig:hdaset} of Appendix B. As with the LDA ice, we see a similar trend where the smallest predicted radii are for ionization, with values that level off at $\sim5.5\pm0.5$ nm. Likewise, following roughly the same trend, but at slightly larger values, the energy deposition radii have values of $\sim4.2$ nm. Also similar to the LDA ice, the total track core radii closely follow the radii of excitation collisions, again with an average radius of $\sim8.0\pm0.5$ nm, which is smaller than the LDA value. 

Equipped now with results for both LDA and HDA ices, we next turn our attention to a consideration of the dependence of the track radii on the material density. Recall from Eq. \eqref{eq:csda_se} in \S1 that, in principle, the track radius is determined by the stopping range of an electron with the average energy of the ejected secondary electrons, resulting in a density dependence of $r_\mathrm{cyl}\propto 1/\rho$. Using the densities we employed here for LDA and HDA ice of 0.94 and 1.1 g cm$^{-3}$, we predict a ratio of  $r_\mathrm{cyl}^\mathrm{HDA}/r_\mathrm{cyl}^\mathrm{LDA}=0.85$. Taking the ratio of the $r_\mathrm{tot}$ values at 6 MeV, we similarly obtain of $r_\mathrm{tot}^\mathrm{HDA}/r_\mathrm{tot}^\mathrm{LDA}=0.85$, thereby nicely recovering the expected density dependence. 

These results are somewhat at odds with the findings of \citet{bringa_electronic_2000}, whose expression for track core radii has an overall density dependence of $r_\mathrm{cyl}\propto n^{1/3}$. Similarly, the peak radii we obtain in our simulations of 9.9 nm and 8.4 nm for LDA and HDA ice, respectively, are roughly double the constant 5 nm value assumed in \citet{leger_desorption_1985} and widely used in astrophysical models. However, as with the LDA results, the track core radii we obtain for HDA ice qualitatively follow the behavior we would expect from Eq. \eqref{eq:csda_se} and Fig. 10 of \citet{rudd_differential_1988}.

\begin{deluxetable}{c|cc|cc|cc|cc}[tb]
\tablecaption{Track radii of energetic protons in amorphous water ice \label{tab:rlda}}
\tablecolumns{9}
\tablewidth{0pt}
\tablehead{
	\colhead{Primary Ion Energy} &
	\multicolumn{2}{c}{$r_\mathrm{tot}$} & 
	\multicolumn{2}{c}{$r_\mathrm{exc}$} & 
	\multicolumn{2}{c}{$r_\mathrm{ion}$} &  
	\multicolumn{2}{c}{$r_\mathrm{energy}$} \\
	\colhead{(MeV)} & 
	\multicolumn{2}{c}{(nm)} & 
	\multicolumn{2}{c}{(nm)} & 
	\multicolumn{2}{c}{(nm)} & 
	\multicolumn{2}{c}{(nm)} \\
	\colhead{} & \colhead{LDA} & \colhead{HDA} & \colhead{LDA} & \colhead{HDA} & \colhead{LDA} & \colhead{HDA} & \colhead{LDA} & \colhead{HDA} 
}
\startdata
0.1   & 3.7 & 3.1 & 3.7 & 3.1 & 0.8 & 0.6 & 1.3 & 1.0 \\
0.2   & 4.3 & 3.5 & 4.3 & 3.5 & 1.5 & 1.2 & 2.0 & 1.7 \\
0.3   & 4.8 & 4.0 & 4.8 & 4.0 & 2.0 & 1.6 & 2.6 & 2.2 \\
0.4   & 5.3 & 4.5 & 5.3 & 4.5 & 2.4 & 2.0 & 3.1 & 2.6 \\
0.5   & 5.8 & 4.8 & 5.8 & 4.8 & 2.7 & 2.3 & 3.5 & 2.9 \\
0.6   & 5.7 & 4.8 & 5.7 & 4.8 & 2.8 & 2.4 & 3.6 & 3.0 \\
0.7   & 6.0 & 5.1 & 6.0 & 5.1 & 3.1 & 2.6 & 3.9 & 3.3 \\
0.8   & 6.3 & 5.3 & 6.3 & 5.3 & 3.4 & 2.8 & 4.2 & 3.5 \\
0.9   & 6.5 & 5.5 & 6.5 & 5.5 & 3.6 & 3.0 & 4.4 & 3.7 \\
1.0   & 6.7 & 5.7 & 6.7 & 5.7 & 3.8 & 3.2 & 4.6 & 3.9 \\
2.0   & 8.2 & 6.9 & 8.2 & 6.9 & 5.4 & 4.5 & 6.1 & 5.1 \\
3.0   & 9.1 & 7.8 & 9.0 & 7.8 & 6.3 & 5.6 & 7.0 & 6.1 \\
4.0   & 9.4 & 8.0 & 9.3 & 7.9 & 6.8 & 5.8 & 7.4 & 6.3 \\
5.0   & 9.7 & 8.8 & 9.6 & 8.7 & 7.3 & 6.7 & 7.8 & 7.1 \\
6.0   & 9.9 & 8.4 & 9.9 & 8.3 & 7.2 & 6.2 & 7.8 & 6.6 \\
7.0   & 9.6 & 8.1 & 9.6 & 8.1 & 7.0 & 5.9 & 7.6 & 6.4 \\
8.0   & 9.4 & 8.2 & 9.4 & 8.2 & 6.8 & 6.0 & 7.4 & 6.5 \\
9.0   & 9.8 & 8.4 & 9.7 & 8.3 & 7.2 & 6.4 & 7.8 & 6.8 \\
10.0  & 9.3 & 8.0 & 9.3 & 8.0 & 6.8 & 5.9 & 7.4 & 6.4 \\
20.0  & 9.3 & 8.1 & 9.3 & 8.0 & 6.7 & 5.9 & 7.3 & 6.4 \\
30.0  & 9.2 & 7.5 & 9.2 & 7.4 & 6.8 & 5.3 & 7.4 & 5.8 \\
40.0  & 8.4 & 7.5 & 8.4 & 7.4 & 5.6 & 5.2 & 6.3 & 5.8 \\
50.0  & 8.3 & 7.3 & 8.2 & 7.2 & 5.4 & 4.8 & 6.1 & 5.4 \\
60.0  & 9.5 & 7.6 & 9.4 & 7.6 & 7.1 & 5.3 & 7.6 & 5.9 \\
70.0  & 8.8 & 7.6 & 8.7 & 7.6 & 6.4 & 4.9 & 6.9 & 5.6 \\
80.0  & 9.0 & 7.9 & 8.9 & 7.9 & 6.2 & 5.8 & 6.9 & 6.4 \\
90.0  & 8.3 & 7.8 & 8.2 & 7.8 & 5.4 & 5.9 & 6.3 & 6.4 \\
100.0 & 8.9 & 8.4 & 8.8 & 8.4 & 5.9 & 6.1 & 6.7 & 6.7 
\enddata
\end{deluxetable}

\begin{figure}
\gridline{
          \fig{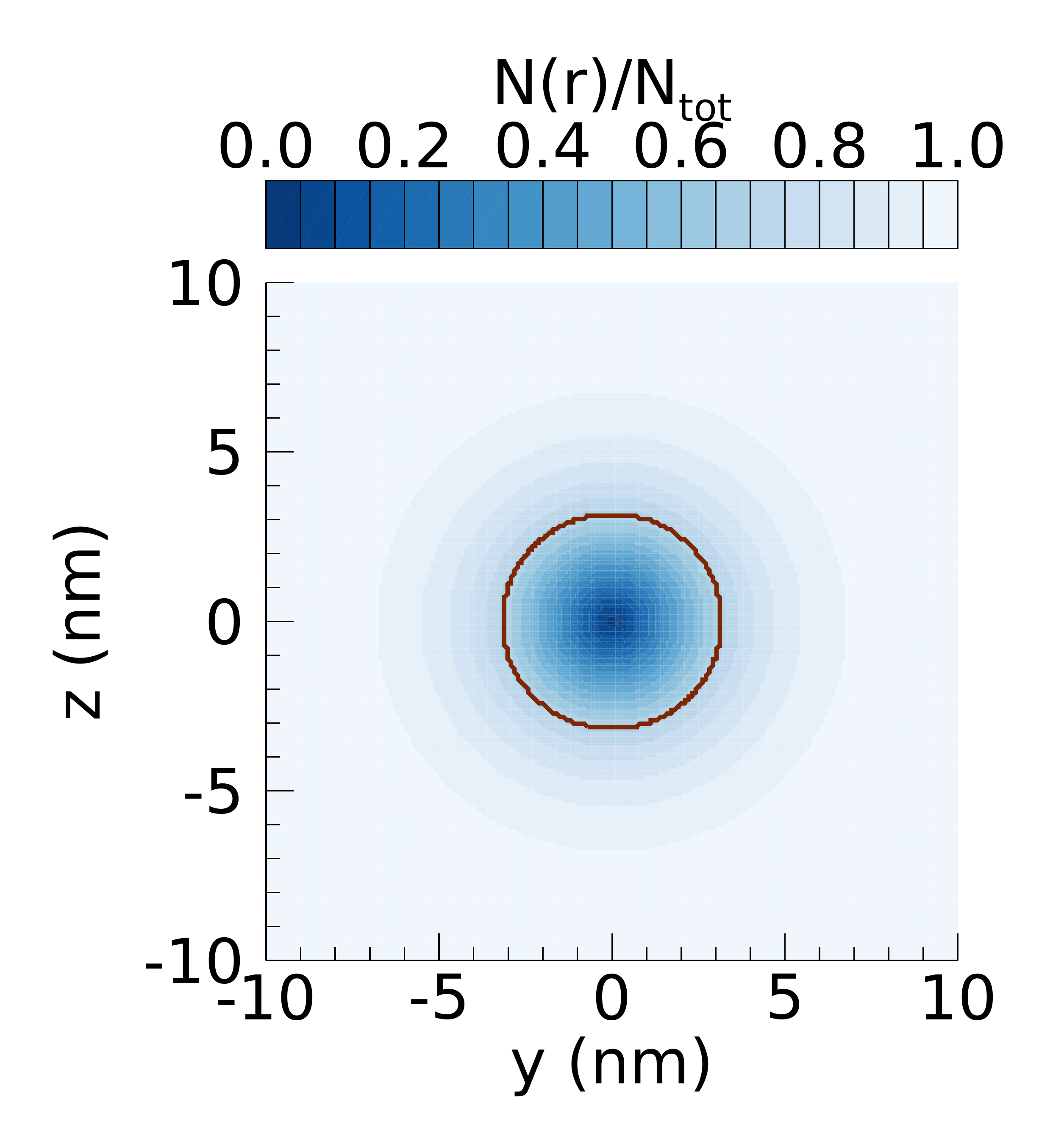}{0.48\textwidth}{(a)}
          \fig{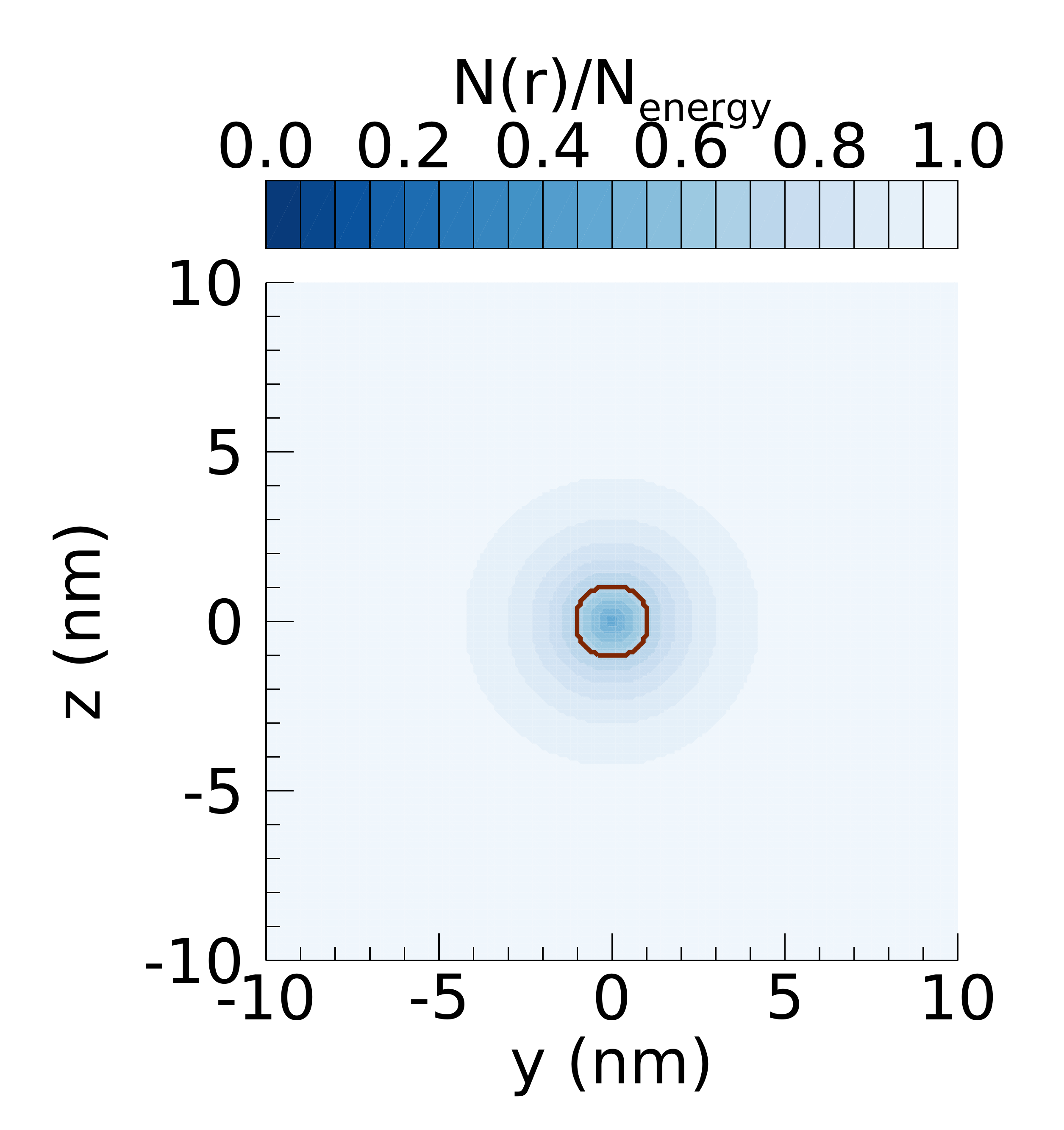}{0.48\textwidth}{(b)}
          }
\gridline{
          \fig{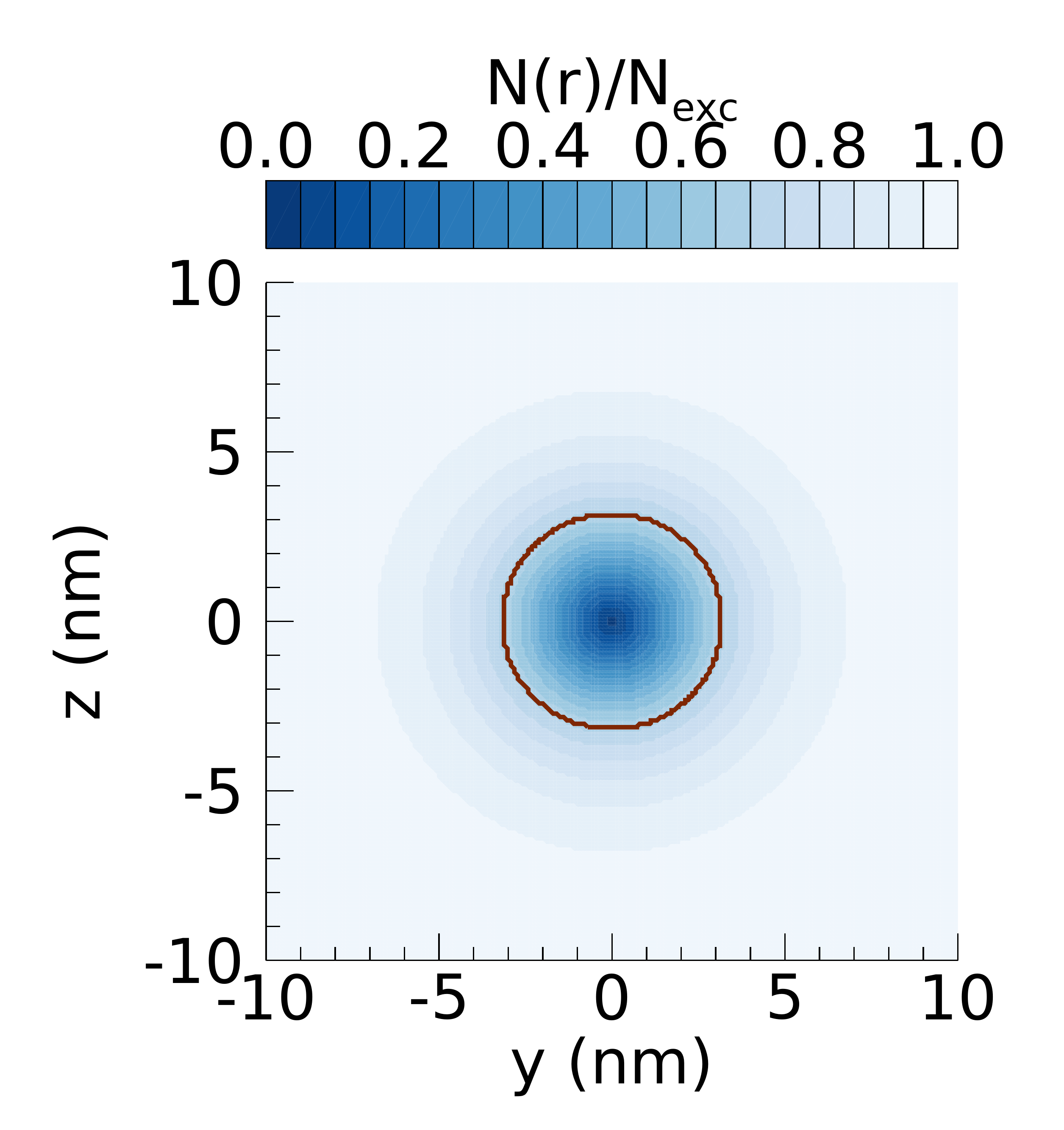}{0.48\textwidth}{(c)}
          \fig{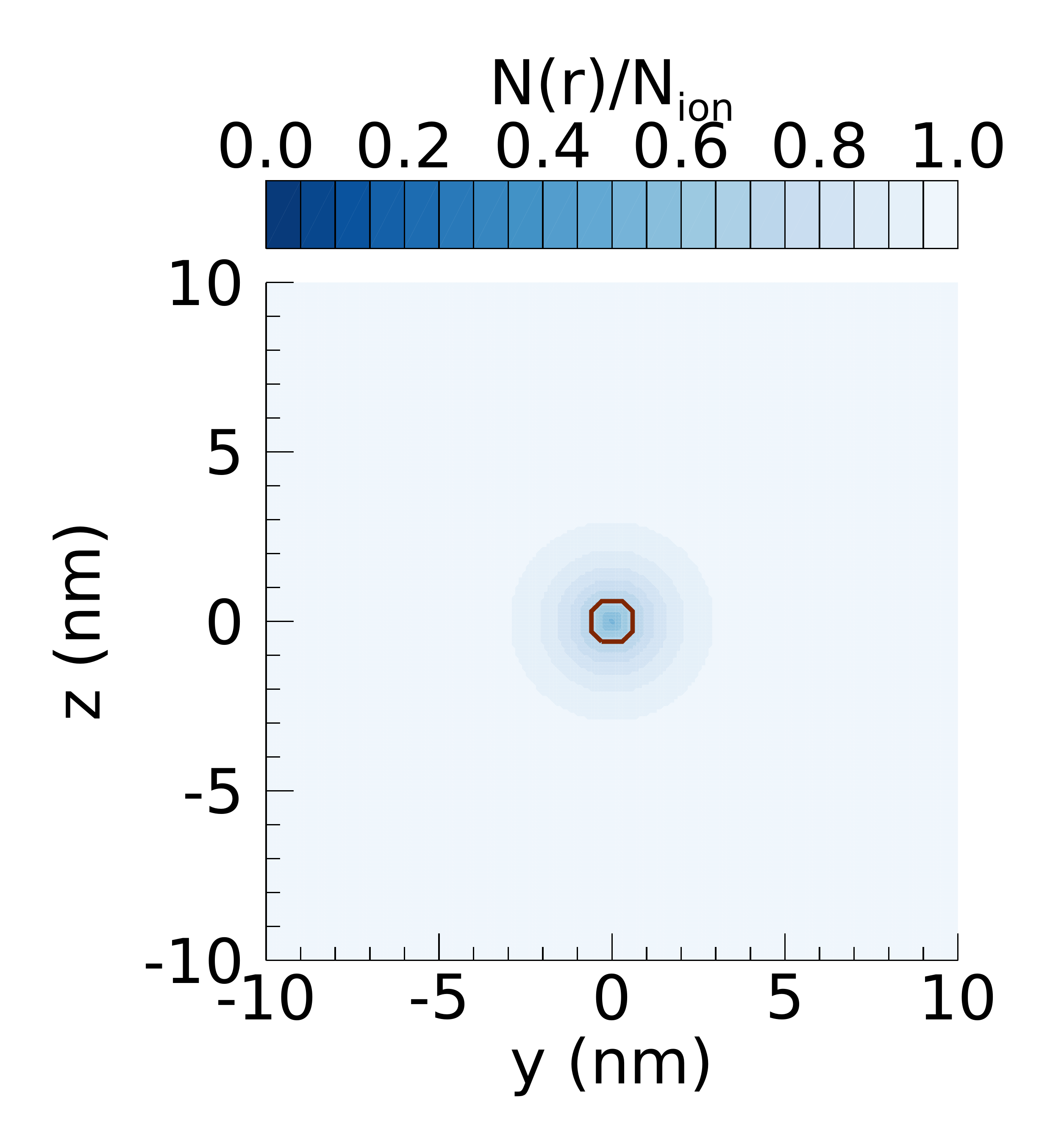}{0.48\textwidth}{(d)}
          }
\caption{The average number of total collisions (a), energy deposition (b), excitation collisions (c), and ionization collisions (d) for a 0.1 MeV \ce{H+} in HDA ice at radius $r$, $N(r)$, over the total value, $N(tot)$. The radius at which $1\sigma$ ($\sim68$\%) of the collisions occur is represented by a red circle. \label{fig:cdf_hda}}
\end{figure}

 \FloatBarrier
\section{Conclusions} \label{sec:conclusions}

Here, we have reported on the results of calculations we have carried out on the track core radii of energetic protons (characteristic of cosmic rays) in both low-density amorphous (LDA) and high-density amorphous (HDA) ices. These calculations were performed using the leading-edge microscopic Monte Carlo toolkit, Geant4-DNA, which, despite its initial focus on radiobiological effects, has also proven to be a useful tool in understanding astrophysically-relevant phenomena due to the commonality of the underlying physics. Our main conclusions are the following:

\begin{enumerate}
    \item track core radii show the weak energy dependence in the range of 0.1 - 6.0 MeV expected from previous calculations of secondary electron energies, but that within the range of 6.0 - 100.0 MeV the radii values stabilize,
    \item the peak track core radii, $r_\mathrm{cyl}$, for LDA and HDA ices are, respectively, 9.9 nm and 8.4 nm - approximately double the radii of 5 nm assumed in \citet{leger_desorption_1985} and to increase somewhat for  incident proton energies below a few MeV, and finally
    \item in agreement with the radii predicted from the electron stopping ranges, our results show a density dependence consistent with $1/\rho$. 
\end{enumerate}

As summarized in \S\ref{sec:introduction}, an accurate knowledge of track core radii is essential for understanding the chemistry and interfacial dynamics of low-temperature irradiated materials. Moreover, a knowledge of $r_\mathrm{cyl}$ is essential for predicting the possible importance of grains explosions \citep{greenberg_exploding_1973,ivlev_impulsive_2015}. Thus, these results should be of great importance in further improving how astrochemical models simulate cosmic ray-irradiated interstellar dust-grain ice mantles \citep{shingledecker_general_2018,shingledecker_cosmic-ray-driven_2018,shingledecker_efficient_2020}. 

Given the foregoing, there are several key implications of our results, related to the larger values of $r_\mathrm{cyl}$ obtained here (for incident protons with energies below a few MeV) compared with, e.g., the constant $r_\mathrm{cyl}=5$ nm assumed in \citet{leger_desorption_1985} and widely adopted in astrochemical models. First, regarding radiation-chemical changes induced by cosmic ray bombardment, our larger radii imply that the volume in which fast reactions involving suprathermal species occur is larger than previously thought, and that a greater fraction of the ice mantle is involved in such chemistry per collision event. Moreover, as noted in the introduction, desorption at the top of cylinder - i.e. at the ice/vacuum interface - is greatly enhanced due to, for instance, the sharp rise in the temperature immediately following cosmic ray bombardment. Thus, the $r_\mathrm{cyl}$ values predicted by this work imply somewhat more efficient desorption via this (and related) mechanisms than what might have been estimated in prior studies.

Finally, we note that this initial study proves the utility of the Geant4-DNA Monte Carlo toolkit in understanding processes of interest in astrochemistry. Given the flexibility of the code, future studies using Geant4-DNA could investigate, e.g., how track radii change with different ice compositions or structures. 

\acknowledgements
    C.N.S. thanks the Alexander von Humboldt Stiftung/Foundation for their generous support. The work by A.V. is supported by the Russian Ministry of Science and Higher Education via the State Assignment Project FEUZ-2020-0038. A.V. is the head of Partner Group of the Max Planck Institute for Extraterrestrial Physics, Garching, at the Ural Federal University, Ekaterinburg, Russia. I. Kyriakou and D. Emfietzoglou acknowledge financial support from the European Space Agency (ESA Contract No. 4000126645/19/NL/BW)

\bibliography{references}{}
\bibliographystyle{aasjournal}

\appendix

\section{Proton Tracks in LDA Ice}

\begin{figure}
\centering
\figurenum{6}
\includegraphics[height=0.8\paperheight]{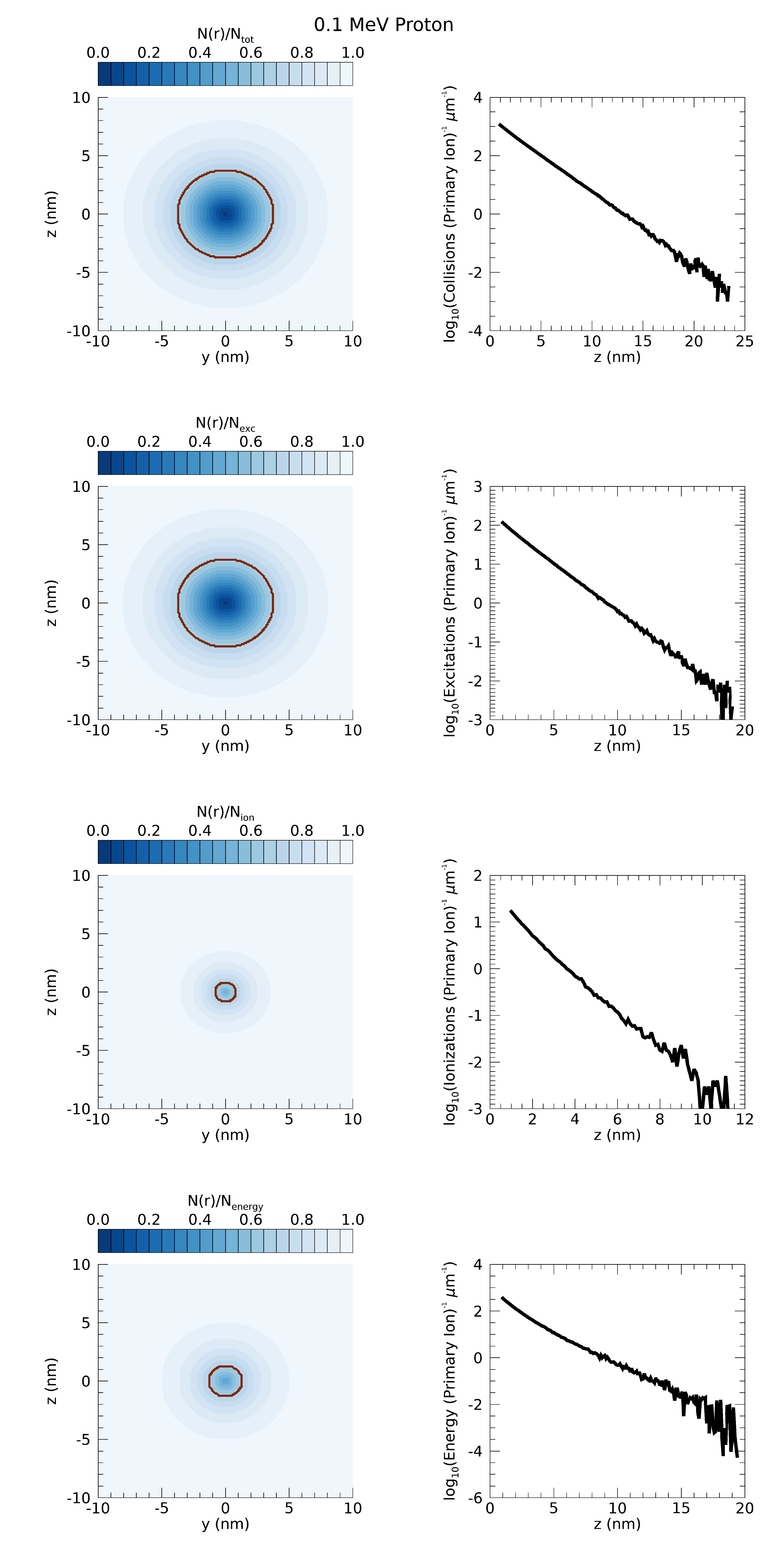}
\caption{1- and 2D representations of the track for 0.1 MeV protons in LDA ice. The complete figure set (28 images) showing the corresponding plots for all energies considered is available in the online journal.}
\label{fig:ldaset}
\end{figure}

In Fig. \ref{fig:ldaset}, we show the 1- and 2D representations of the total track, as well as the several sub-tracks considered here, namely, those showing specifically the excitation or ionization collisions, as well as the average energy deposition.

\figsetstart
\figsetnum{6}
\figsettitle{Track Images for LDA Ice}

\figsetgrpstart
\figsetgrpnum{6.1}
\figsetgrptitle{2D (left column) and 1D (right column) representations of the total collisions (row 1), excitation collisions (row 2), ionization collisions (row 3), and energy deposition (row 4) for a 0.1 MeV Proton in LDA ice}
\figsetplot{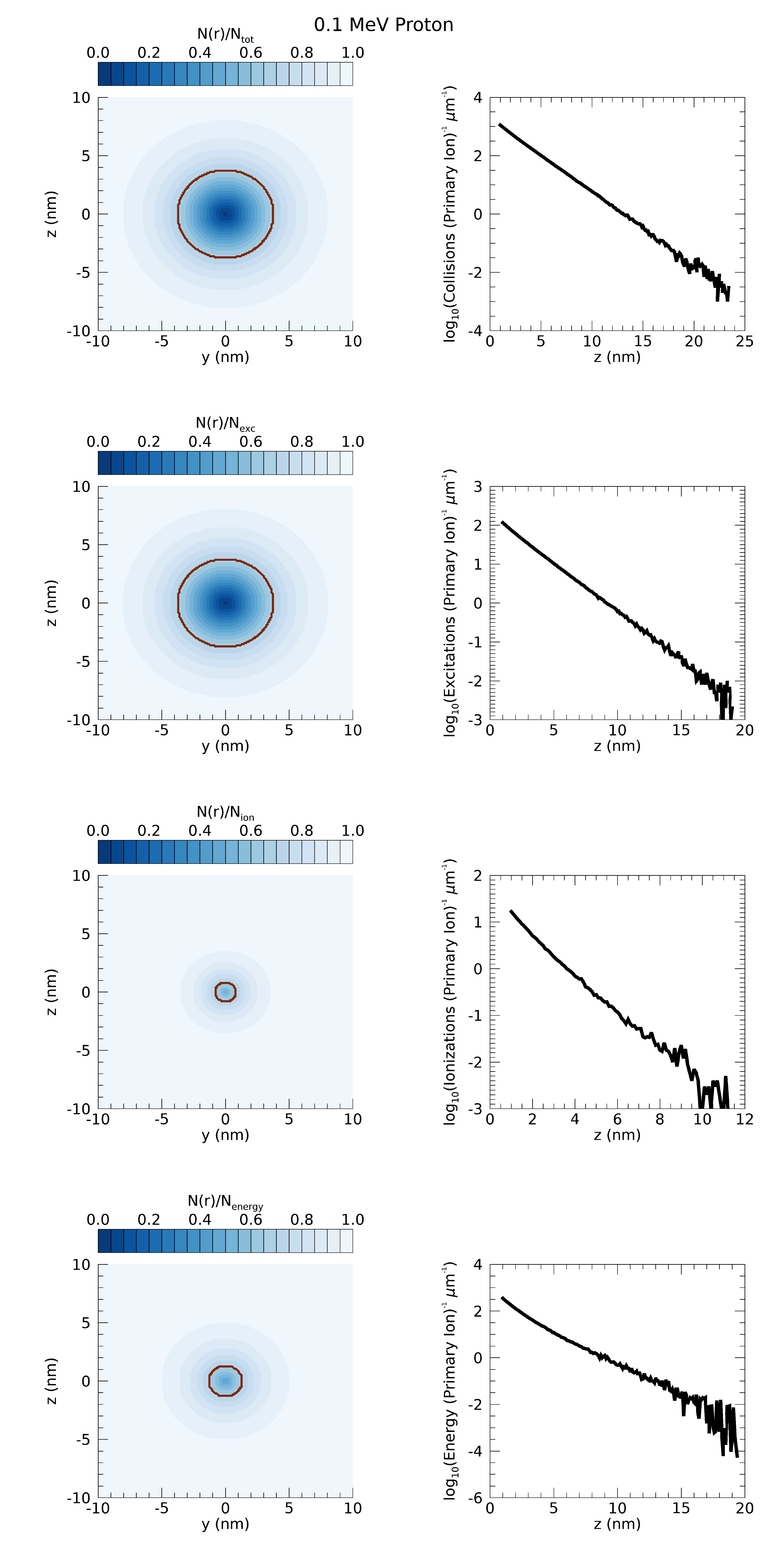}
\figsetgrpnote{1- and 2D representations of the track for protons in LDA ice. The complete figure set (28 images) is available in the online journal.}
\figsetgrpend

\figsetend

\section{Proton Tracks in HDA Ice}

\begin{figure}
\centering
\figurenum{7}
\includegraphics[height=0.8\paperheight]{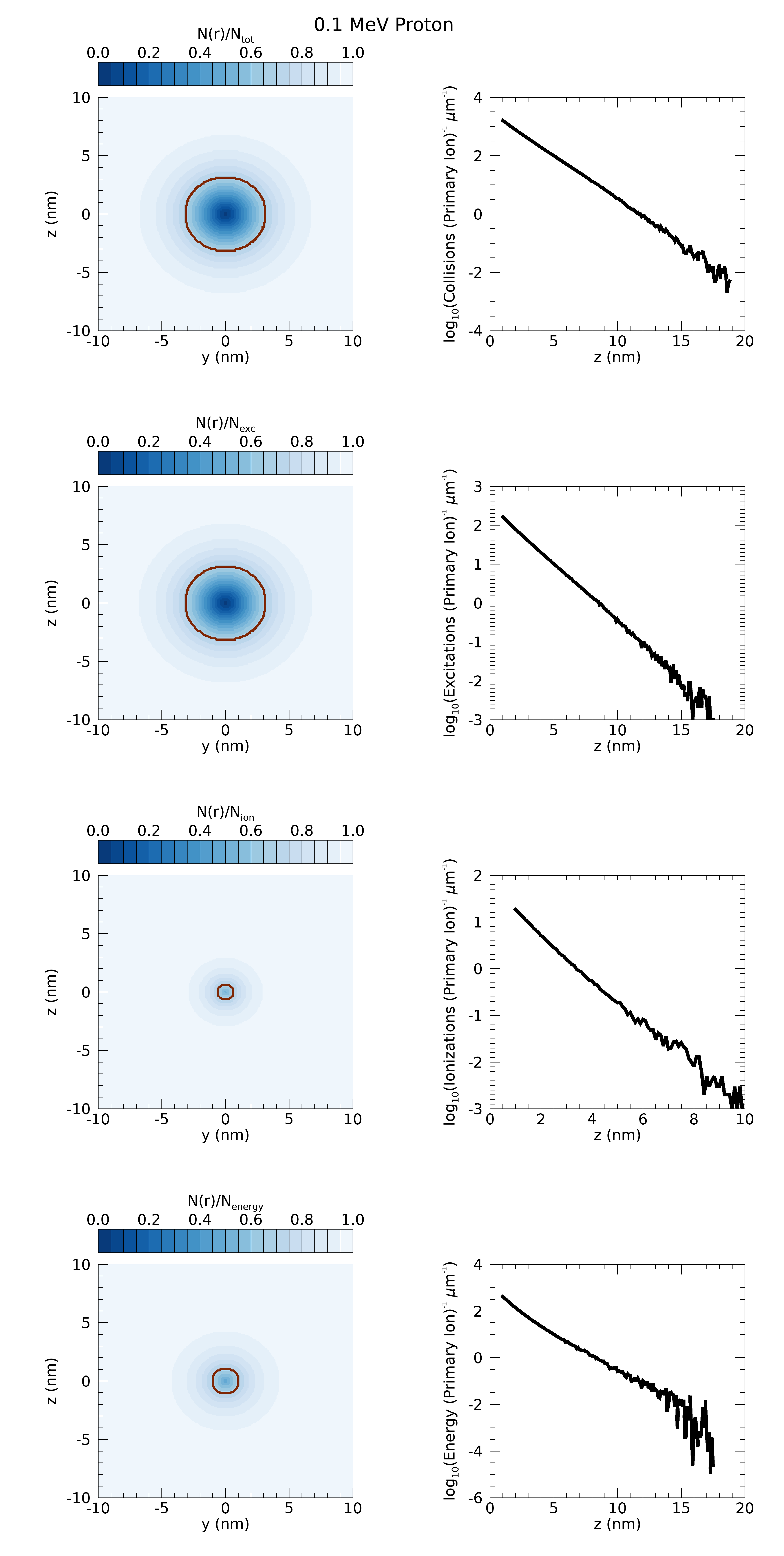}
\caption{1- and 2D representations of the track for 0.1 MeV protons in HDA ice. The complete figure set (28 images) showing the corresponding plots for all energies considered is available in the online journal.}
\label{fig:hdaset}
\end{figure}

In Fig. \ref{fig:hdaset}, we show the 1- and 2D representations of the total track, as well as the several sub-tracks considered here, namely, those showing specifically the excitation or ionization collisions, as well as the average energy deposition.

\figsetstart
\figsetnum{7}
\figsettitle{Track Images for HDA Ice}

\figsetgrpstart
\figsetgrpnum{7.1}
\figsetgrptitle{2D (left column) and 1D (right column) representations of the total collisions (row 1), excitation collisions (row 2), ionization collisions (row 3), and energy deposition (row 4) for a 0.1 MeV Proton in HDA ice}
\figsetplot{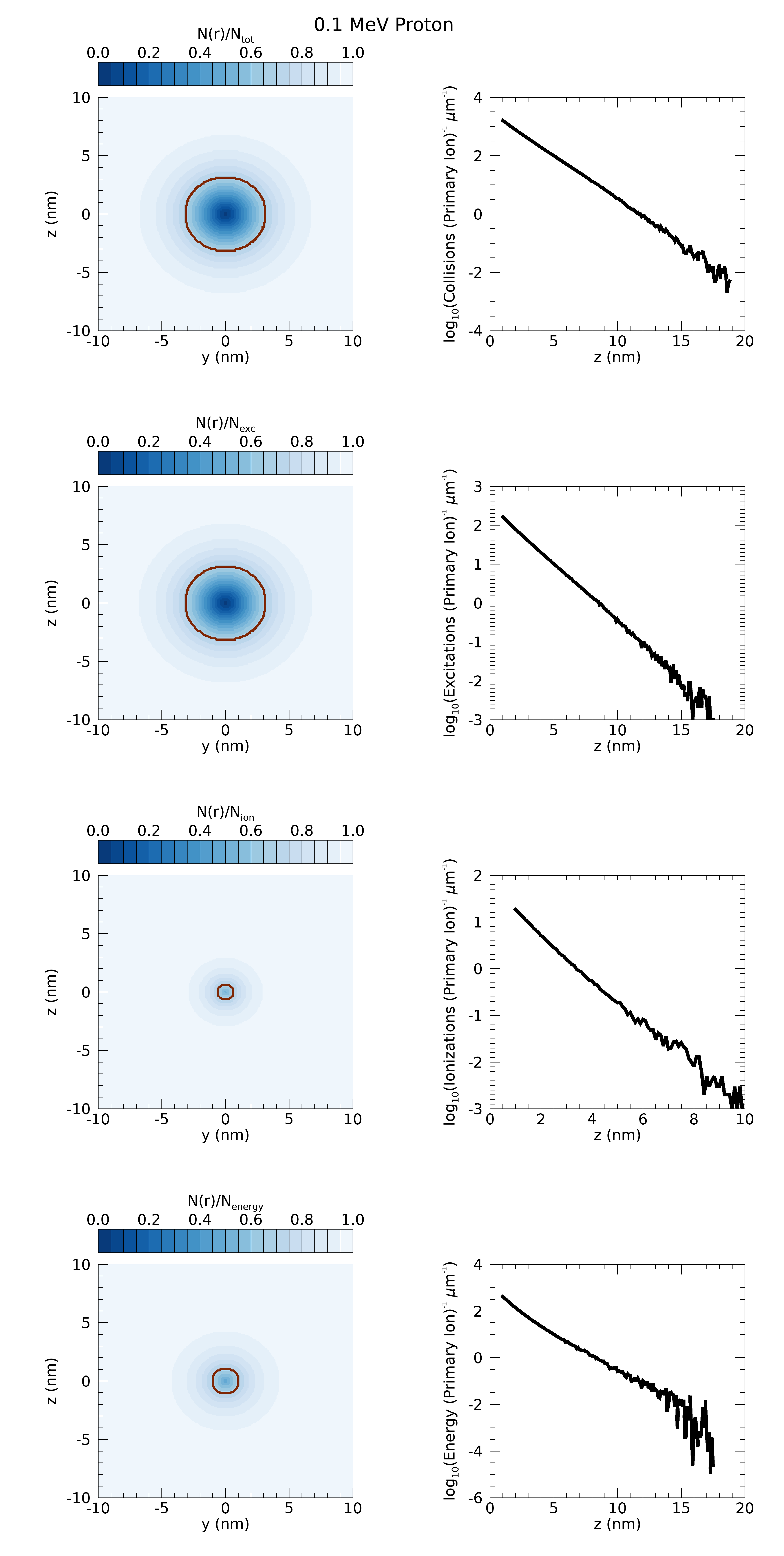}
\figsetgrpnote{1- and 2D representations of the track for protons in HDA ice. The complete figure set (28 images) is available in the online journal.}
\figsetgrpend

\figsetend

\end{document}